\documentclass[]{tandf2}

\usepackage{epstopdf}% To incorporate .eps illustrations using PDFLaTeX, etc.
\usepackage{subfigure}% Support for small, `sub' figures and tables

\usepackage[numbers,sort&compress]{natbib}% Citation support using natbib.sty
\bibpunct[, ]{[}{]}{,}{n}{,}{,}% Citation support using natbib.sty
\renewcommand\bibfont{\fontsize{10}{12}\selectfont}% Bibliography support using natbib.sty
\makeatletter
\def\NAT@def@citea{\def\@citea{\NAT@separator}}% Suppress spaces between citations using natbib.sty
\makeatother

\theoremstyle{plain}% Theorem-like structures provided by amsthm.sty

\theoremstyle{definition}

\theoremstyle{remark}

\usepackage{amsmath}
\usepackage[numbers]{natbib}
\usepackage{lineno}
\modulolinenumbers[5]
\usepackage[]{setspace}
\usepackage{epsfig}
\usepackage{color}
\usepackage{url}
\usepackage{colordvi}
\usepackage{ifthen}
\usepackage{txfonts}
\usepackage{multicol}
\usepackage{multirow}
\usepackage{graphicx}
\usepackage{bm}

\newcommand{\meanN}{\langle N\rangle}
\newcommand{\meanT}{\langle T\rangle}

\newcommand{\meanIb}{\langle I_b\rangle}
\newcommand{\meanI}{\langle I\rangle}
\newcommand{\meankappa}{\langle \kappa\rangle}

\begin{document}

%\articletype{ARTICLE TEMPLATE}

\title{Multipoint Radiation Induced Ignition of Dust Explosions: Turbulent Clustering of Particles and Increased Transparency}

\author{
\name{M. Liberman\textsuperscript{a}$^{\ast}$\thanks{$^\ast$Corresponding author. Email: mliber@nordita.org}, N. Kleeorin\textsuperscript{b,a}, I. Rogachevskii\textsuperscript{b,a} and N. E. L.~Haugen\textsuperscript{c,d}}
\affil{
 \textsuperscript{a}Nordita, Stockholm University and Royal Institute of Technology, Roslagstullsbacken 23, 10691, Stockholm, Sweden;\\
 \textsuperscript{b}Department of Mechanical Engineering, Ben-Gurion University
 of the Negev, P. O. Box 653, Beer-Sheva 84105, Israel;\\
 \textsuperscript{c}Department of Energy and Process Engineering,
  Norwegian University of Science and Technology
  Kolbj{\o}rn Hejes vei 1B, NO-7491 Trondheim, Norway;\\
  \textsuperscript{d}SINTEF Energy Research, N-7465 Trondheim, Norway}
}

\maketitle

\begin{abstract}
It is known that unconfined dust explosions consist of a relatively weak primary (turbulent) deflagrations followed by a devastating secondary explosion. The secondary explosion may propagate with a speed of up to 1000 m/s producing overpressures of over 8-10 atm. Since detonation is the only established theory that allows a rapid burning producing a high pressure that can be sustained in open areas, the generally accepted view was that the mechanism explaining the high rate of combustion in dust explosions is deflagration to detonation transition. In the present work we propose a theoretical substantiation of the alternative propagation mechanism explaining origin of the secondary explosion producing the high speeds of combustion and high overpressures in unconfined dust explosions. We show that clustering of dust particles in a turbulent flow gives rise to a significant increase of the thermal radiation absorption length ahead of the advancing flame front. This effect ensures that clusters of dust particles are exposed to and heated by the radiation from hot combustion products of large gaseous explosions sufficiently long time to become multi-point ignition kernels in a large volume ahead of the advancing flame front. The ignition times of fuel-air mixture by the radiatively heated clusters of particles is considerably reduced compared to the ignition time by the isolated particle. The radiation-induced multi-point ignitions of a large volume of fuel-air ahead of the primary flame efficiently increase the total flame area, giving rise to the secondary explosion, which results in high rates of combustion and overpressures required to account for the observed level of overpressures and damages in unconfined dust explosions, such as e.g. the 2005 Buncefield explosion and several vapor cloud explosions of severity similar to that of the Buncefield incident.
\end{abstract}


%\preprint{NORDITA-2017-104}

\section{Introduction}

Understanding the origin and mechanism of dust explosions is essential for minimizing the dust explosion hazard in many industrial processes, to understand their causes, prevention and mitigation methods. The danger of dust explosions is a permanent threat in various industries handling combustible or inert powders of fine particles
\citep{Eckhoff(2003),Eckhoff(2009),Abbasi(2007),Yuan(2015),Atkinson(2015),Atkinson(2017),Li(2016)}.
Methane-air systems are life-threatening mixtures, particularly in underground coal mines. Methane from coal mines initiated explosions, and the explosions are later escalated by coal dusts \citep{Kundu(2016)}. Experimental studies \citep{Rockwell(2013)} show that particles increase the turbulent burning velocity because of the increase in the turbulence level by particle-turbulence interaction. In cases with smaller particle sizes and a relatively larger dust concentrations ($>50-70$ g/m$^3$) the turbulent burning velocity increases considerably larger compared with larger particle sizes and lower dust concentration ranges. Explosions initiated by these mixtures have destroyed infrastructure in mines and have taken thousands of lives in the past. Faraday and Lyell, who in 1845 analyzed \citep{Faraday(1845)} one of the most destructive coal mine explosions, the one in the Haswell coal mine, had been the first who pointed to the likely key role of dust particles. In the real world, the types and nature of dust explosions are diverse. Devastating dust explosions can occur in a system of the combustible or inert particles suspended in fuel-air.
The catastrophic Aluminium particles dust explosions is another example of devastating explosions, occurred on August 2014 in a large industrial plant in Kunshan, China), where 75 people were killed and 185 were injured \citep{Li(2016)}.

	Dust explosions can be either partly or fully confined (e.g. explosions of pulverized bio-mass plants), or unconfined. In majority of the large-scale vapor cloud explosions (VCEs) there was clear forensic evidence that a severe explosion had propagated into open uncongested areas. This was a feature of all of the large vapor cloud incidents for which detailed primary evidence was available, e.g., the 2005 Buncefield fuel storage depot explosion), and several VCEs of severity similar to that of the Buncefield, which occurred during past decade \citep{Atkinson(2015),Atkinson(2017),Herbert(2013),Atkinson(2011)}.
Unconfined dust explosions and VCEs start with a primary accidentally ignited flame in the deflagration regime. In majority of dust explosions the primary flame becomes turbulent due to obstacles in the flame propagation path, which can significantly enhance the flame propagation speed.

Historically, large-scale chemical explosions that involve a fuel (methane, propane or butane) mixed with the atmosphere were considered as unconfined dust explosions. Admittedly, a large-scale coal mine explosions share traits with the large vapor cloud incidents and may be considered as either partly or almost fully unconfined dust explosions. 
Large-scale and unconfined dust explosions and VCEs start with a primary accidentally ignited flame in the deflagration regime. In majority of dust explosions, the primary flame becomes turbulent due to obstacles in the flame propagation path, which can significantly enhance the flame propagation speed. A dust explosion occurs when the dust particles are dispersed in the fuel-air ahead of the advancing flame.
In real explosions it is common for the pressure waves from the primary ignited flame to disturb dust on the ground ahead of the flame front, so a dust explosion occurs when the dust particles are entrained and 
dispersed in the fuel-air. However, despite considerable efforts over more than 100 years, the mechanism governing flame propagation in dust explosions that cause an extremely high combustion speeds and overpressures, required to account for the levels of observed damage, still remains a major unresolved issue. Currently most prevention and safety methods rely on empirical correlations. Although the technology for preventing dust explosions has progressed considerably, recurring accidents in the mining industries and fuel storages (VCEs) call for a fundamentally improved understanding of the mechanism of the flame propagation and the origin of high overpressures in these explosions.

\section{Is DDT the only possible way to explain unconfined dust explosions?}

The conventional flame propagation mechanisms in the deflagration regime cannot cause  high overpressures and explain the observed level of damages such as, e.g. in the Buncefield incident. Since detonation is the only established theory that allows sufficiently rapid burning while producing a high pressure that can be sustained in open areas, this led many researchers to conjecture (see e.g. \citep{Bradley(2012),Herbert(2013),Pekalski(2015)}) that deflagration to detonation transition (DDT) following turbulent flame acceleration is the mechanism that explains the high rate of combustion and overpressures and the observed damages in dust explosions.

	Already from the early experimental studies of the deflagration to detonation transition in ducts \citep{Shchelkin(1940),Shchelkin(1965),Urtiew(1966),Oppenheim(1973),Moen(1980),Ciccarelli(2008)} it was known that the flame accelerates more rapidly if it passes through an array of turbulence-generating baffles. Turbulence acts as a perturbing force, increasing the flame surface and the overall burning rate. Channels with rough walls and obstacles are often used to study DDT, since in this case the run-up distance is better controlled
\citep{Teodorczyk(2009),Ciccarelli(2013),Wang(2015),Wang(2016)}. This was presumably the reason  why the first attempts to explain DDT were heavily based on the assumption that DDT might occur only if the flame becomes turbulent, and that turbulence is the primary reason for the flame acceleration. At the same time, already at the beginning of DDT studies, Ya. B. Zeldovich \citep{Zeldovich(1947)} has convincingly shown that stretching of the flame front due to the interaction with a nonuniform upstream flow velocity field in the boundary layer is the main process of the flame acceleration, while turbulence plays a supplementary role. Stretching of the flame front by the sheared flow is much stronger mechanism of the flame acceleration, than the flame front wrinkling caused by turbulence. Indeed, DDT was observed experimentally in a smooth tube for different highly reactive mixtures, e.g. hydrogen-oxygen or ethylene–oxygen
\citep{Kuznetsov(2005)}. It was shown theoretically and by using numerical simulations
\citep{Liberman(2010),Ivanov(2011)}, that transition to detonation occurs,
if the flame sufficiently accelerates in the flow behind the leading shock wave, so that it can catch up the leading shock and the positive feedback between the shock and the rate of chemical reaction of the flame may lead to transition to detonation.

	Deflagration-to-detonation transition in unconfined systems is more problematic. The process is scale dependent, therefore the investigations in a small laboratory scales cannot reveal this phenomenon. The run-up distance for transition to detonation increases with the transverse tube diameter. For a channel with repeated obstacles the run-up distance is of the order or larger than (10-20) of the tube diameters. The experimental evidence for the increased turbulent flame speeds, were reported in large-scale congested space for various highly reactive fuels in air (\citep{Pekalski(2015)}, and references therein). According to experimental studies \citep{Kuznetsov(2002)}, the critical tube diameter, for which transition to detonation in methane-air can occur, should be not less than 9.5 m, and the transition can occur at the run up distance exceeding 100-150 m.

In the recent theoretical and numerical models possible mechanism of DDT in truly unconfined systems is attributed to the flame acceleration induced by the flame front wrinkling caused by
the flame hydrodynamic (DL) instability \citep{Kagan(2017),Koksharov(2017)}
or turbulence \citep{Poludnenko(2011)}.
On the other hand, numerical simulations \citep{Poludnenko(2011)} demonstrated that possible DDT in unconfined media, caused by the flame interactions with turbulence, is feasible only for a very large turbulent velocities.
In general, theoretical models whose purpose is to explain possible mechanisms of DDT in unconfined systems led to a deeper understanding of the effects of shocks, turbulence, and other mechanisms of flame acceleration on creating the conditions in which DDT could occur. Nonetheless, even though these studies advanced basic understanding, they opened up more questions than it answered. It is also possible that in dust explosions the additional flame acceleration due to radiative preheating can promote DDT in unconfined systems, though this would require a special investigation, which is outside the scope of the present study.

The possibility that DDT occurs in vapor cloud explosions in industrial accidents has generally only been recognized in the experiments for highly reactive fuels, such as hydrogen and ethylene, and the corresponding models has been validated against a range of experimental data, obtained from a laboratory-scale experiments.
Applying it to industrial accidents
is taking the models beyond its validation range, where they cannot be used as a predictive tool. In particular, in the case of the Buncefield and similar VCE incidents, the extent and density of congestion are substantially less than that required for DDT.

Several severe explosions extending to the whole cloud
has been investigated in the years since the Buncefield incident. Forensic evidences and detailed reviews of many such incidents have shown serious discrepancies between a presumption, that the overpressures and damages of a variety of objects were caused by detonation and what has been observed at most VCE incidents \citep{Atkinson(2015),Atkinson(2017),Atkinson(2011)}. A detailed analysis of physical damages and data available from CCTV cameras, led to conclusions
\citep{Atkinson(2011),Atkinson(2017)} that scenarios based on a detonation \citep{Herbert(2013),Bradley(2012)} and the type of the observed damages
are not consistent with that would occur in a detonation, and that \citep{Atkinson(2011)}: ''the combustion in Buncefield was unsteady (episodic), with periods of rapid flame advance (producing high local overpressures) being punctuated by pauses. Overall the average rate of flame progress was sub-sonic ($\sim$ 150 m/s)''.

In most large scale dust explosion accidents,
a series of explosions consisting of a primary weak explosion, followed by a devastating secondary explosion, has been reported \citep{Eckhoff(2003),Eckhoff(2009),Yuan(2015),Atkinson(2015),Atkinson(2017)}. While the hazardous effect of the primary explosion is relatively small, the secondary explosions propagate with a speed of up to 1000 m/s, producing overpressures of in the range of 8--10 bar. The main goal of the present study is to clarify a mechanism, governing flame propagation in unconfined dust explosions. We suggest an alternative mechanism, explaining episodic flame propagation mode, featuring periods of high rates of combustion and overpressures, punctuated by slow flame propagation in unconfined dust explosions.

\section{Effect of the radiative preheating}

The role of radiation in dust flame propagation is an important issue that was not
sufficiently examined so far. The impact of thermal radiation emanated by the advancing flame in large dust explosions can be expected with a range of possible outcomes. Particles suspended in an initially flammable gas mixture heated by thermal radiation, 
in turn, heat the surrounding gas and 
can enhance the flame speed and cause ignition of the surrounding fuel-air mixture. This effect can play an important role in the large dust explosions and VCEs. In unconfined dust explosions the radiative flux emitted from the advancing flame into the reactants is significantly enhanced by the increased emissivity of the large volume of burned products with hot particles. According to \cite{Nathan(2012),Hadjipanayis(2015)} 
the forward radiative flux, emitted by hot combustion products in large gaseous explosions, is close to the blackbody radiation at stoichiometric flame temperatures (2200 -- 2500 K). Recent advances in diagnostics methods have shown that radiation can make a decisive contribution to the overall energy transport, structure and velocity of flames \citep{Nathan(2012),Hadjipanayis(2015),Bidabadi(2013),Gao(2015)}. The magnitude of the radiative flux can be expected to be of significant importance. The experimentally measured thermal radiation, caused by dust explosions, and reported values of a maximum surface emissive power are in the range 140 -- 556 kW/m$^2$ \citep{Holbrow(2000),Roberts(2000)}.

	Let us consider the flame propagating in a gaseous fuel-air mixture with evenly dispersed particles. The thermal radiation, emitted from the flame surface, is absorbed and reemitted by the particles ahead of the flame with heat being transferred from the particles to the surrounding gas by thermal conduction. The intensity of the radiant flux decreases exponentially on the scale of the order of the radiation absorption length,
$L_a=1/\meankappa \approx (2\rho_p/3\rho_d)\, d_p$, where $\meankappa = \sigma_p \meanN$ is the mean particle absorption coefficient, $\sigma_p \approx\pi d_p^2/4$ is the particle absorption
cross section, $d_p$ is the particle size (diameter), $\rho_p$ is the material density of the dust particles, $\rho_d$  is the spatial particle density and  $\meanN$ is the mean number density of evenly dispersed particles. The radiation absorption length for the dust cloud mass densities, $\rho_d = 0.01 - 0.05$ kg/m$^3$ typical for dust explosions, is in the range of 1 -- 10 cm for the microns size particles $d_p = 1 - 20 \, \mu$m. For the evenly dispersed particles the maximum increase in the temperature of the gas mixture immediately ahead of the flame front can be estimated as
\begin{eqnarray}
\Delta T_p \approx \frac{0.63 \, \sigma T_b^4}{U_{\rm f}(\rho_p
\, c_{\rm p} + \rho_g \, c_{\rm v})},
\label{A1}
\end{eqnarray}
where $U_{\rm f}$ is the flame velocity, $\sigma T_b^4$ is the
black-body radiative flux, $\rho_g$ is the mass density of gaseous mixture, $c_{\rm p}$ and
$c_{\rm v}$ are specific heats of particles and gas phase, respectively \citep{Liberman(2015)}. The increase in temperature of the radiatively heated particles and gas mixture ahead of the flame depends on the time, $L_a/U_{\rm f}$,
required for the advancing flame to catch-up  the radiatively heated particles ahead of  the flame front.
As smaller the flame speed, as longer is the time of the radiative heating, and as larger is
the temperature increase of the unreacted gas mixture ahead of the flame.

\begin{figure}
\vspace*{1mm} \centering
\includegraphics[width=8.0cm]{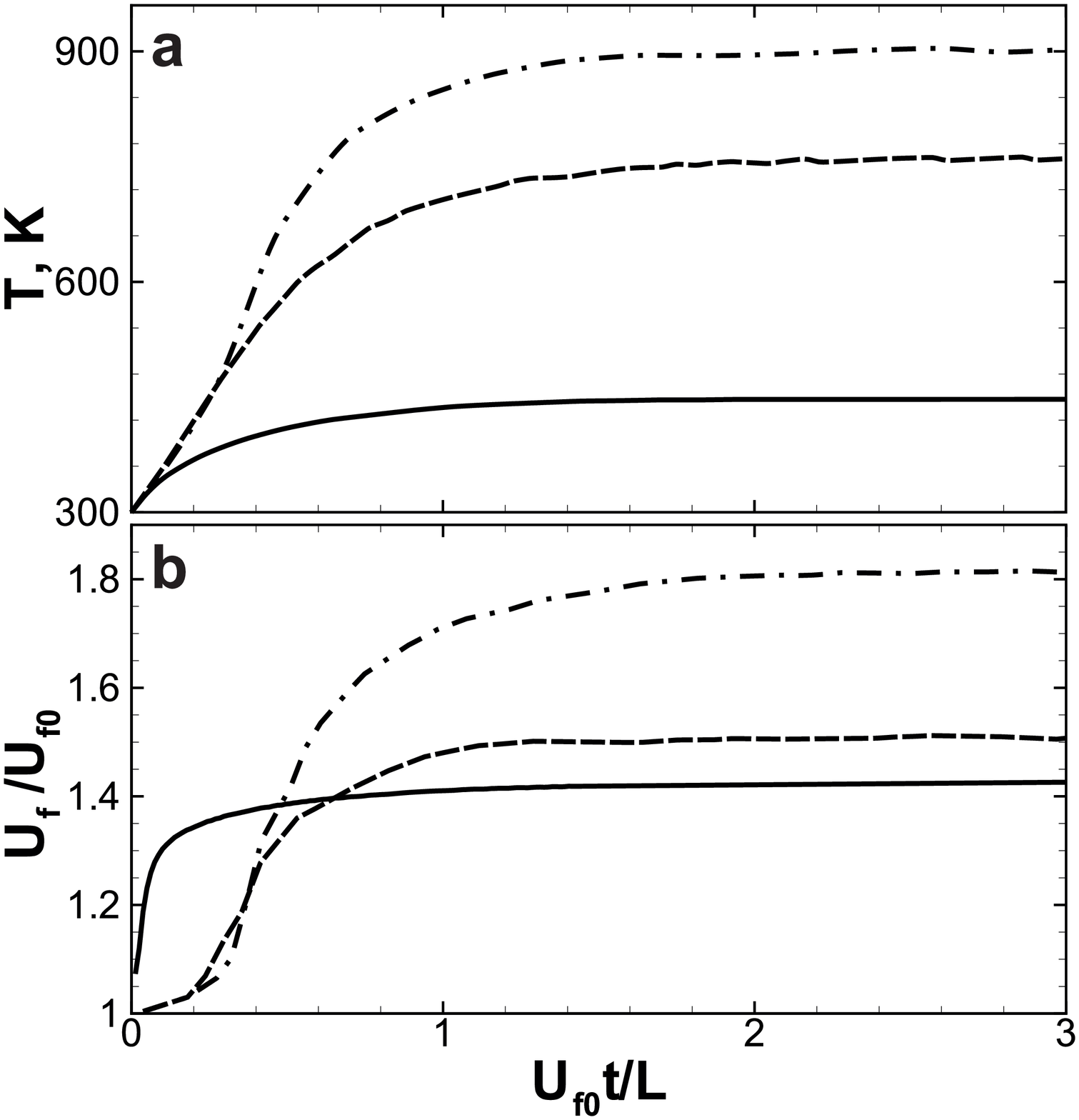}
\caption{\label{Fig1}
Time evolution of the temperature at 2 mm ahead of the flame front (a) and the flame velocity increase (b) for different initial pressures of the H$_2$/O$_2$ mixture: solid line $P_0 =1$ atm, dashed line $P_0 =0.3$ atm, dashed-doted line $P_0 =0.2$ atm. The flame velocities are normalized on the laminar flame velocity $U_{{\rm f}0}$ in pure gas mixture. The radiation absorption length, $L_a=1$ cm, time is in units $L_a/U_{{\rm f}0}$.}
\end{figure}

\begin{figure}
\vspace*{1mm} \centering
\includegraphics[width=8.0cm]{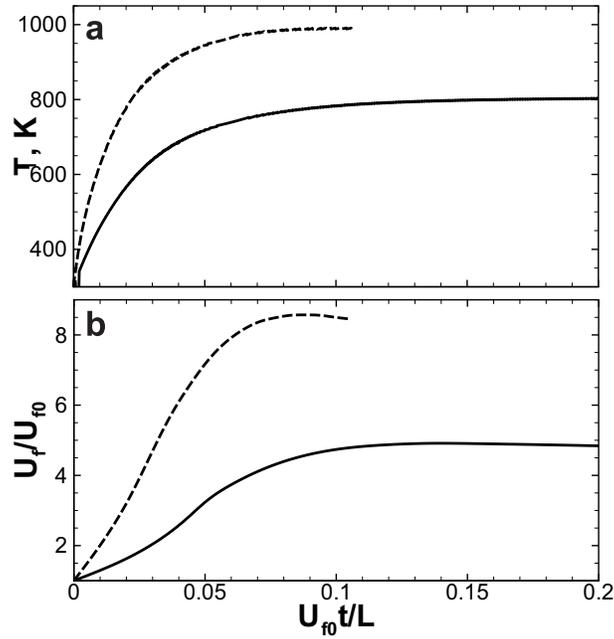}
\caption{\label{Fig2}
 Time evolution of the temperature 2 mm ahead of the flame front (a) and the flame velocity increase (b) during radiative heating for methane-air mixture laden with micron-size particles; solid line $P_0 =1$ atm,  dashed line  $P_0 =0.5$ atm. The flame velocities are normalized on the laminar flame velocity $U_{{\rm f}0}$ in pure gas mixture. The radiation absorption length, $L_a=1$ cm, time is in units $L_a/U_{{\rm f}0}$.
}
\end{figure}

Figures~\ref{Fig1} and~\ref{Fig2} show representative examples of the time evolution of the gas temperature and the corresponding increase of the flame velocity computed for the hydrogen-oxygen (Figure~\ref{Fig1}), and methane-air (Figure~\ref{Fig2}) flames propagating through the particle laden mixtures with uniformly distributed particles for the radiation absorption length, $L_a=1$ cm. The numerical simulations were performed in a way similar as it was done in \citep{Liberman(2015),Ivanov(2015)} for a planar flame propagating in a particle-laden combustible gas mixture. The governing equations for the gaseous phase are the one-dimensional, time-dependent, multispecies reactive Navier-Stokes equations including the effects of compressibility, molecular diffusion, thermal conduction, viscosity and detailed chemical kinetics for the reactive species, production of radicals, energy release and heat transfer between the particles and the gas. The dynamics of solid particles is considered in continuous hydrodynamic approximation. The interaction between particles is assumed negligibly small for a small volumetric concentration of particles, so that only the Stokes force between the particle and gaseous phase is taken into account
\citep{Liberman(2015),Ivanov(2015)}. The high temperature of the particle-laden mixture behind the flame front is considered optically-thick so that the radiation energy flux emitted from the flame is assumed to be equal to the blackbody radiative heat source. The initial velocity of the flame was taken to be normal laminar flame velocity with the temperature of the mixture
ahead of the flame being 300 K.
Time evolution in numerical simulations was studied up to establishing of the steady state for radiatively heated mixture ahead of the flame and the flame speed.

	Basically, the temperature of the radiatively heated unreacted gas ahead of the flame can exceed the ignition threshold for a small enough flame velocity. However, this is not the case for flames in a real combustible mixtures. As a flame propagates through a dust cloud of evenly dispersed particles, it consumes the unburned fuel before the temperature of particles and the gas mixture ahead of the flame will have risen up to the ignition level. Although for slow flames, such as the methane-air flame ($U_{\rm f0} \approx 0.45$ m/s), the radiatively heated mixture results in a noticeable increase of the flame velocity,
yet the radiative heat transfer probably cannot become a dominant process of the heat transfer at least for chemical flames.

	Possibility of ignition depends on the level of thermal radiation
absorbed by the particles, which must be heated to sufficiently high temperatures. In the case of evenly dispersed particles radiation is absorbed by particles in the nearest layers ahead of the flame front, and its intensity decreases exponentially.
Therefore, there is no enough time for particles to be heated up to the level sufficient for ignition of surrounding gas mixture.
The situation is completely different when particles are non-uniformly dispersed, for example, in the form of the optically thick dust layer separated from ahead of the flame front by the gaseous gap with sufficiently small concentration of particles in the gap. In this case the gap between the flame and the layer is transparent for radiation, and particles in the layer are sufficiently long time exposed to and heated by the radiation from the flame to become ignition kernels. It was shown \citep{Ivanov(2015)} that such optically thick layer of particles can ignite a new combustion wave in the surrounding fuel-air mixture ahead of the flame front before it is reached by the advancing flame.

	Ignition of dust clouds by combustible particles, heated up by intense electromagnetic radiation from an infrared laser source, has been demonstrated experimentally by Proust \citep{Proust(2006)}. Detailed experimental investigations of ignition of butane-air mixtures by particles, heated up by intense electromagnetic radiation from an infrared laser, were carried out in \citep{Beyrau(2013),Beyrau(2015)}. The ignition was demonstrated for a range of different carbon particles, as well as for non-reactive particles, such as silicon carbide. For non-combustible particles, the particle temperature necessary for ignition, was 1200 $\pm$ 200 K. Similar times of ignition were obtained also for combustible particles, though it was suggested that the particle material reactivity may influence the ignition temperatures \citep{Beyrau(2015)}. The particle size was found to have a significant impact on the timescale of ignition with shorter ignition times for finer particles.

	The plausibility of formation of local ignition kernels ahead of the flame can play an important role in unconfined dust explosions and vapor cloud incidents. Moore and Weinberg \citep{Moore(1981)} hypothesized that fibrous particles a few millimeters across can be heated and ignited sufficiently quickly by the forward radiation from hot combustion products. This in turn could lead to multi-point ignition of a gas mixture ahead of the flame front resulting in efficiently anomalously high rate of combustion. They have shown also \citep{Moore(1983)} that loose agglomerates of fine fibers can induce ignition of a surrounding fuel-air
when they are heated by radiation (400 kW/m$^2$) from a ${\rm CO}_2$ laser. It has been concluded that particles, suspended in a fuel-air mixture, can cause ignition of the mixture, when they are strongly irradiated, e.g. by radiation from a large flame front or explosion. Li and Lindstedt \citep{Li(2017)} explored another possibility of inducing secondary explosions via radiative ignition of dust particles trapped in the corners around obstacles ahead of the advancing flame, which is similar to the scenario proposed in \cite{Atkinson(2011)}.

	However, it should be noted that in order for a secondary explosion producing a strong pressure or shock wave to occur, a multi-point ignition of a large volume of fuel-air mixture ahead of the advancing flame must be ignited by the ignition kernels of particles heated by radiation from hot combustion products. The fast multi-point ignition of a large volume of fuel-air ahead of the flame front ensures that the pressure from the ignited fuel-air mixture would rise faster than it can be equalized by sound waves. This requires the penetration length of radiation, $L_{\rm eff}$, to be so large, that the particles are sufficiently long time exposed and heated by the radiation, to become ignition kernels in a large volume even far ahead of the flame front, i.e.
\begin{eqnarray}
L_{\rm eff} \gg c_{\rm s} \, \tau_{\rm ign} ,
\label{A2}
\end{eqnarray}
where $c_{\rm s}$ is the speed of sound in the flow ahead of the flame and $\tau_{\rm ign}$ is the timescales of fuel-air ignition by the radiatively heated ignition kernels.

	In practical dust explosion incidents, the plausibility of formation of ignition kernels ahead of the flame is determined by the absorption (penetration) length of radiation and ignition timescales. This is because particles well ahead of the flame front have to be exposed to and sufficiently long time heated by thermal radiation in order to ignite surrounding mixture. We will show that clustering of particles in a turbulent flow gives rise to a strong, up to 2--3 order of magnitude, increase of the radiation penetration length. The time of ignition of a fuel-air mixture by clusters of particles is also considerably reduced, as compared with the ignition time for the isolated radiatively heated particle. The proposed mechanism of the multi-point radiation-induced ignitions due to the turbulent clustering of particles, may explain the occurrence of the secondary explosion regardless of whether the particles are combustible or unreactive.

\section{Effect of turbulent clustering of dust particles on the radiation heat transfer}

In turbulent flows ahead of the primary flame, dust particles with material density, that is much larger than the fuel-air gas density, assemble in small clusters with sizes about several Kolmogorov viscous scales. Here, $\ell_\eta=\ell_0/{\rm Re}^{3/4} \approx 0.1 - 10$ mm
is the Kolmogorov viscous scale, ${\rm Re} = u_0 \ell_0/\nu$ is the Reynolds number based on the integral scale $\ell_0$ of turbulence and the characteristic turbulent velocity $u_0$ at the integral scale.
The turbulent eddies, acting as small centrifuges, push the particles to the regions between the eddies, where the pressure fluctuations are maximum and the vorticity intensity is minimum. Therefore, suspended small particles in a turbulent flow tend to assemble in clusters with much higher particle number densities than the mean particle number density. This effect, known as inertial clustering, has been investigated in a number of analytical \citep{Elperin(1996),Balkovsky(2001),Elperin(2002),Elperin(2007)}, numerical \citep{Bec(2007),Toschi(2009)}, and experimental \citep{Warhaft(2009),Xu(2008),Balachandar(2010),Saw(2014)} studies.

	Our analytical studies \citep{Elperin(2013),Elperin(2015)} and laboratory experiments \citep{Eidelman(2010)} have shown that the particle clustering is significantly enhanced in the presence of a mean temperature gradient, so the turbulence is temperature stratified and the turbulent heat flux is not zero. This causes correlations between fluctuations of fluid temperature and velocity, and non-zero correlations between fluctuations of pressure and fluid velocity. This enhances the particle clustering in the regions of maximum pressure fluctuations. As a result, the particle concentration in clusters rises by a few orders of magnitude, as compared to the mean concentration of evenly dispersed particles \citep{Elperin(2013),Elperin(2015)}.

	It is known, that spatial inhomogeneities with scales larger than the wavelength of the radiation, may give rise to an increase of the radiation absorption length \citep{Kliorin(1989),Kravtsov(1993),Apresyan(1996)}. The effect of an increase in the radiation transmission through gas-particle mixtures caused by the inertial particle clustering in non-stratified turbulence was studied using  a Monte Carlo model in different particle distributions \citep{Farbar(2016),Frankel(2016)}. In the case of the non-stratified turbulence and for relatively small Reynolds numbers both effects: the particle clustering and the increase of the radiation absorption length are not very pronounced. On the contrary, in the case of a dust explosion, there are many reasons for the formation of an inhomogeneous temperature distribution and temperature gradients in the turbulent flow produced by pressure waves ahead of the primary flame. We will show that the clustering of particles in the temperature stratified turbulence ahead of the primary flame gives rise to a strong increase of the radiation penetration length. The effect ensures that the inequality~(\ref{A2}) is satisfied because clusters of particles even far ahead of the primary flame are sufficiently long time exposed and heated by the radiation from the primary flame to become ignition kernels.

\subsection{Equation for the mean radiation intensity}  	

	To investigate how clustering of particles affects the radiation penetration length, and how this effect depends on the turbulence parameters, we consider a turbulent flow with suspended particles exposed by a radiative flux. The radiative transfer equation for the intensity of radiation, $I({\bm r},\hat{\bm s})$, in the two-phase flow reads
(see, e.g., \citep{Zeldovich(1966),Howell(2010)}):
\begin{eqnarray}
\left(\hat{\bm s} {\bm \cdot} {\bm \nabla}\right)
I({\bm r},\hat{\bm s}) &=&  - \left[\kappa_g({\bm r})
+ \kappa_p({\bm r})+\kappa_s({\bm r})\right]\, I({\bm r},\hat{\bm s})
+ \kappa_g \, I_{b,g} + \kappa_p I_{b,p}
\nonumber\\
&& + {\kappa_s \over 4 \pi} \int \phi({\bm r},\hat{\bm s},\hat{\bm s}')
\, I({\bm r},\hat{\bm s}') \,d\Omega ,
\label{A3}
\end{eqnarray}
where $\kappa_g({\bm r})$ and $\kappa_p({\bm r})$ are the absorption coefficients
of gas and particles, respectively, $\kappa_s({\bm r})$ is the particle
scattering coefficient, $\phi({\bm r},\hat{\bm s},\hat{\bm s}')$ is the scattering phase function, $I_{b,g}({\bm r})$ and $I_{b,p}({\bm r})$ are the black-body radiation intensities
for gas and particles, ${\bm r}$ is the position vector, respectively,
$\hat{\bm s}={\bm k}/k$ is the unit vector in the direction of radiation, ${\bm k}$ is the wave vector, $(k, \theta, \varphi)$ are the spherical coordinates in ${\bm k}$ space, and
$\,d\Omega=\sin \theta \,d\theta \,d\varphi$.
Taking into account that the scattering and absorption cross sections for gases at normal conditions are very small, the contribution from the gas phase is negligible in comparison with that of particles. We also take into account that for $\pi d_p \sqrt{|\epsilon|}/\lambda_{\rm rad} \gg 1$, the scattering coefficient is negligibly small compared with the particle absorption coefficient \citep{Howell(2010)}, where $\epsilon$ is the dielectric permeability of the dust particles, $\lambda_{\rm rad}$ is the radiation wavelength and $d_{p}$ is the diameter
of the dust particles.  Therefore, Eq.~(\ref{A3}) is reduced to
\begin{eqnarray}
&& \left(\hat{\bm s} {\bm \cdot} {\bm \nabla}\right)
I({\bm r},\hat{\bm s}) = - \kappa({\bm r}) \left(I({\bm r}) - I_b\right) ,
\label{A4}
\end{eqnarray}
where $\kappa \equiv\kappa_p({\bm r})$ and $I_b\equiv I_{b,p}$ that depends on the local temperature, and we omitted subscript ‘p’.

	To derive equation for the effective radiation absorption coefficient in the presence of particle clustering in a turbulent flow, we use a mean-field approach. In the framework of this approach, all quantities are decomposed into the mean and fluctuating parts:  $I=\meanI + I'$, $I_b=\meanIb + I'_b$ and $\kappa=\meankappa + \kappa'$. The fluctuating parts, $I', I'_b, \kappa'$, have zero mean values, and angular brackets
denote averaging over an ensemble of fluctuations.

The radiation absorption length for evenly dispersed particles is $L_a=1/\meankappa$, where $\meankappa = \sigma_a \meanN$ is the mean particle absorption coefficient, $\sigma_a\approx\pi d_p^2/4$ and $\meanN$  is the mean number density of evenly dispersed particles.
The particle absorption coefficient is $\kappa=n \, \sigma_a$, so that the fluctuations of the absorption coefficient are $\kappa' = n' \, \sigma_a= n' \, \meankappa/\meanN$.
Averaging Eq.~(\ref{A4}) over the ensemble of the particle number density fluctuations, we obtain the equation for the mean irradiation intensity $\langle I({\bm r},\hat{\bm s})\rangle$:
\begin{eqnarray}
&& \left(\hat{\bm s} {\bm \cdot} {\bm \nabla}\right)
\langle I({\bm r},\hat{\bm s})\rangle = - \meankappa
\left(\meanI - \meanIb\right) - \langle \kappa' \, I' \rangle
+ \langle \kappa' \, I'_b \rangle .
\label{A5}
\end{eqnarray}
Subtracting Eq.~(\ref{A5}) from Eq.~(\ref{A4}), we obtain the equation for fluctuations of $I'$:
\begin{eqnarray}
&& \left(\hat{\bm s} {\bm \cdot} {\bm \nabla} + \meankappa
+ \kappa' \right) I'({\bm r},\hat{\bm s}) = I_{\rm source} ,
\label{A6}
\end{eqnarray}
where $s={\bm r} {\bm \cdot} \hat{\bm s}$, and the source term is
\begin{eqnarray}
&& I_{\rm source}= - \kappa' \, \left(\meanI - \meanIb\right)
+ \langle \kappa' \, I' \rangle +
\left(\meankappa + \kappa'\right)I'_b
 - \langle \kappa' \, I'_b \rangle  .
\label{A7}
\end{eqnarray}
The solution of Eq.~(\ref{A6}) is
\begin{eqnarray}
&& I'({\bm r},\hat{\bm s}) = \int_{-\infty}^{\infty}
I_{\rm source} \, \exp\left[-\left|\int_{s'}^{s}
\left[\meankappa+ \kappa'(s'') \right] \,ds'' \right| \right]\,ds' .
\label{A8}
\end{eqnarray}
Expanding the exponent, $\exp\left[-\int_{s'}^{s} \kappa'(s'') \,ds''\right]$, in Eq.~(\ref{A8}) in Taylor series over a small parameter, $\kappa' \ell_\eta {\rm Sc}^{-1/2} \ll 1$ one obtains,
\begin{eqnarray}
\exp\left[-\int_{s'}^{s} \kappa'(s'') \,ds''\right] = 1 - \int_{s'}^{s} \kappa'(s'') \,ds'' + {\rm O}\left(\kappa'^2\right) ,
\label{A8a}
\end{eqnarray}
so that Eq.~(\ref{A8}) reads:
\begin{eqnarray}
&& I'({\bm r},\hat{\bm s}) = \int_{-\infty}^{\infty}
I_{\rm source} \, \exp\left(-\meankappa |s-s'|\right) \,
\left(1 - \int_{s'}^{s} \kappa'(s'') \,ds'' \right) \,ds'
+ {\rm O}\left(\kappa'^2\right) ,
\nonumber\\
\label{A8b}
\end{eqnarray}
where ${\rm Sc}=\nu/D$ is the Schmidt number, $\nu$ is the kinematic viscosity, $D$ is the coefficient of the Brownian diffusion of particles.
Multiplying Eq.~(\ref{A8b}) by $\kappa'$ and averaging over the ensemble of fluctuations, we obtain expression for the one-point correlation function $\langle \kappa' \, I' \rangle$:
\begin{eqnarray}
&& \langle \kappa' \, I' \rangle \, \left[1 + \int_{-\infty}^{\infty}
\left(\int_{s'}^{s} \langle\kappa'(s) \kappa'(s'') \rangle \,ds'' \right) \, \exp\left(-\meankappa |s-s'|\right) \, \,ds' \right]
\nonumber\\
&& \quad = - \Big(\meanI - \meanIb\Big) \int_{-\infty}^{\infty}
\langle\kappa'(s) \kappa'(s') \rangle \,
\exp\left(-\meankappa |s-s'|\right) \, \,ds' .
\label{A8c}
\end{eqnarray}
To derive Eq.~(\ref{A8c}), we used the quasi-linear approach \citep{Liberman(2017)}. In the framework of this approach, we neglected in Eq.~(\ref{A8c}) small third and higher moments in fluctuations of $\kappa'$, because a small parameter , $\kappa' \ell_\eta {\rm Sc}^{-1/2} \ll 1$. We also neglected a small correlation between the particle number density fluctuations and the gas temperature fluctuations, which implies that $\langle \kappa' \, I'_b \rangle = 0$.

Equation~(\ref{A8c}) can be rewritten as
\begin{eqnarray}
\langle \kappa' \, I' \rangle  = - \meankappa
\Big(\meanI - \meanIb\Big) \, {2 \beta J_1 \over 1 + 2 \beta J_2} ,
\label{A9}
\end{eqnarray}
where the values $J_1$ and $J_2$  in Eq.~(\ref{A9}) are given by integrals:
\begin{eqnarray}
&& J_1= \int_{0}^{\infty} \Phi(Z) \exp(-\beta Z) \,dZ ,
\label{A10}\\
&& J_2= \beta \int_{0}^{\infty} \left(\int_{0}^{Z} \Phi(Z') \,dZ' \right)
\, \exp(-\beta Z) \,dZ ,
\label{A11}
\end{eqnarray}
$\Phi(t,{\bm R}) = \langle n'(t,{\bm x}) \, n'(t,{\bm x} + {\bm R})\rangle$ is the two-point correlation function of the particle number density fluctuations,
$\beta=\meankappa \, \ell_D$ and $\ell_D =a \, \ell_\eta / {\rm Sc}^{1/2}$, with the numerical factor $a \gg 1$.
Substituting the correlation function $\langle \kappa' \, I' \rangle$, determined by Eq.~(\ref{A9}) into Eq.~(\ref{A5}), we obtain the equation for the mean radiation intensity:
\begin{eqnarray}
\left(\hat{\bm s} {\bm \cdot} {\bm \nabla}\right)
\langle I({\bm r},\hat{\bm s}) \rangle = - \kappa_{\rm eff} \,
\Big(\meanI - \meanIb\Big) ,
\label{A12}
\end{eqnarray}
where the effective absorption coefficient $\kappa_{\rm eff}$, which takes into account the particle clustering in a temperature stratified turbulence, is
\begin{eqnarray}
\kappa_{\rm eff} = \meankappa \, \left(1 - {2 \beta J_1 \over 1 + 2 \beta J_2}\right) .
\label{A13}
\end{eqnarray}

\subsection{Effect of particle clustering}

To calculate integrals $J_1$ and $J_2$, we use the normalized two-point correlation function of the particle number density fluctuations, $\Phi(R)=\langle n'({\bm r}) \,
n'({\bm r}+{\bm R}) \rangle / \meanN^2$. For a temperature stratified turbulence, the two-point correlation function which accounts for particle clustering in the temperature stratified turbulence, was derived in \cite{Elperin(2013)}:
$\Phi(R) = (n_{\rm cl} / \meanN)^2$, for $0\leq R\leq \ell_D$, and
\begin{eqnarray}
\Phi(R) = \left({n_{\rm cl} \over \meanN}\right)^2 \,
\left({R \over \ell_D}\right)^{-\mu} \cos\left[\alpha \,
\ln \left({R \over \ell_D}\right)\right],
 \label{A15}
\end{eqnarray}
for $\ell_D \leq R\leq \infty$.
This correlation function determines the maximum increase of the particle number density inside a cluster, where $R={\bm R} {\bm \cdot} \hat{\bm s}$,
\begin{eqnarray}
\mu &=& {1 \over 2(1+3\sigma_{_{T}})} \left[3 - \sigma_{_{T}} + {20 \sigma_v (1+\sigma_{_{T}})\over 1+\sigma_v}\right],
\label{B1}\\
\alpha &=& {3\pi (1+\sigma_{_{T}})\over (1+3\sigma_{_{T}}) \, \ln {\rm Sc}} .
\label{B2}
\end{eqnarray}
Here $\sigma_{_{T}} = \left(\sigma_{_{T0}}^2 + \sigma_v^2\right)^{1/2}$
is the degree of compressibility of the turbulent diffusion tensor, $\sigma_{_{T0}} = \sigma_{_{T}}({\rm St}=0)$, and
$\sigma_v$ is the degree of compressibility of the particle velocity field:
\begin{eqnarray}
\sigma_v &=& {8 {\rm St}^2 \, K^2 \over 3(1 + b \, {\rm St}^2 \, K^2)} ,
\; \; K = \left[1+ {\rm Re} \,\left({L_\ast \, {\bm \nabla} \meanT \over \meanT}
\right)^2\right]^{1/2} ,
 \label{A18}
\end{eqnarray}
${\rm Re} = \ell_0 u_0 /\nu$ is the Reynolds number,
$\meanT$ is the mean fluid temperature,
the length $L_\ast = c_{\rm s}^2 \tau_\eta^{3/2}/9 \nu^{1/2}$,
and a constant $b$ determines the value of the parameter
$\sigma_v$ in the limit of ${\rm St}^2 \, K^2 \gg 1$.
To derive Eqs.~(\ref{A15})--(\ref{A18}), we used Eqs. (42)--(46)
in \cite{Elperin(2013)} and Eqs. (17)--(18) in \cite{Elperin(2015)}.

The particle inertia is characterised by
the Stokes number, ${\rm St} =\tau_p/{\tau}_\eta$,
where $\tau_p=m_p/(3 \pi \rho \, \nu d_p)$ is the particle Stokes
time, $\tau_\eta= \tau_0 /{\rm Re}^{1/2}$ is
the Kolmogorov time, $\tau_0=\ell_0/u_0$ and $u_0$
are the characteristic time and velocity at the turbulent
integral scale, $\ell_0$, respectively.
The size of a cluster is given by $\ell_D =a \, \ell_\eta / {\rm Sc}^{1/2}$.
The value of the numerical factor $a \approx 5 - 10$ corresponds to the values obtained in
the laboratory experiments \citep{Eidelman(2010)} and the atmospheric measurements \citep{Elperin(2015),Sofiev(2009)}.
The parameter $b$ enters in the expression for the degree of compressibility
of particle velocity field, $\sigma_v$.
In the limit of large temperature gradients (${\rm St}^2 \, K^2 \gg 1$),
the value $\sigma_v = 8/3b$.  Since the degree of compressibility
of particle velocity field $\sigma_v \leq 1$, the coefficient $b \geq 8/3$.
On the other hand, when $b$ is very large ($b \gg 1$), $\sigma_v \to 0$, i.e.,
the role of  stratification of turbulence or particle inertia is negligible,
which implies the absence of the turbulent clustering.
Therefore, it is reasonable to choose the value
of $b$ in the range $5 \leq b \leq 10$.
Our analysis shows that the final result for the effective radiation penetration length, $L_{\rm eff}$, is not very sensitive to the particular
value of $b$ (see below).

In the analytical study of the radiative transfer, we neglect the third and higher
moments in $\kappa'$ in the correlation function $\langle \kappa' \, I' \rangle$,
due to a small parameter in the theory, $\kappa' \ell_\eta
{\rm Sc}^{-1/2} \ll 1$. Since for inertial particles, ${\rm Sc} = \nu/D \gg 1$,
this condition is satisfied.
At the same time, in the study of turbulent clustering of particles, we have taken into
account higher moments in the particle number density fluctuations,
which contributes to the effective  radiation penetration length.
The higher moments in $\kappa'$ (i.e., in fluctuations of particle number density
$n'$) are only important within the particle clusters, while outside the clusters
the higher moments in $\kappa'$ are much smaller than the second moments.

\begin{figure}
\vspace*{1mm} \centering
\includegraphics[width=11.0cm]{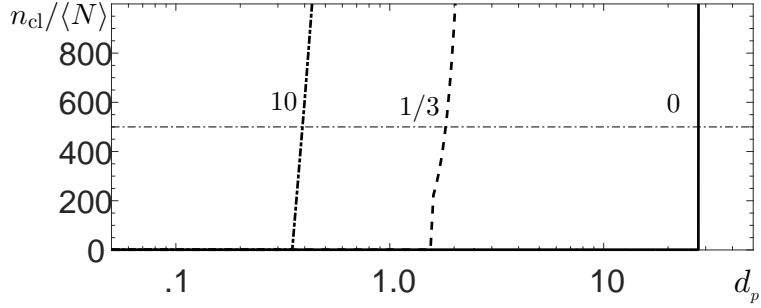}
\caption{\label{Fig3}
The particle number density, $n_{\rm cl} / \meanN$, within the cluster normalized by the mean
particle number density versus the particle diameter $d_p$ for different temperature gradients $|{\bm \nabla}\meanT|=$ 10 K/m (dashed-dotted); 1/3 K/m (dashed); 0 K/m (solid).
The inclined lines are related to the first saturation mechanism due to the particle depletion effect. The horizontal line corresponds to the second saturation mechanism caused by the momentum coupling of particles and turbulent flow with the mass loading parameter $m_p n_{\rm cl}/ \rho_g \approx 1/2$.
}
\end{figure}

	The growth of particles number density inside the cluster is saturated by nonlinear effects. The particle number density inside the cluster can be restricted by depletion of particles in the surrounding flow caused by their accumulation inside the cluster.
It was shown in \cite{Elperin(2013)} that the maximum number density of particles attained
inside the cluster and caused by the particle depletion effect is given by:
\begin{eqnarray}
{n_{\rm cl} \over \meanN} = \left(1 + {e \, \mu
\over \pi} \, {\rm Sc}^{\mu/2} \, \ln {\rm
Sc}\right)^{1/2} .
 \label{B10}
\end{eqnarray}
Another effect, that restricts the growth of the particle number density inside the cluster, is related to a momentum coupling of particles and turbulent flow, when kinetic energies of particles and turbulence become close. Numerical modeling shows that the latter effect becomes important when the mass loading parameter (the ratio of the particle spatial density inside the cluster to the fluid density), $m_p n_{\rm cl}/ \rho_g \sim 1$.
In Figure~\ref{Fig3} we show the particle number density, $n_{\rm cl} / \meanN$, within the cluster normalized by the mean particle number density versus the particle diameter $d_p$ for different temperature gradients $|{\bm \nabla}\meanT|=$ 10 K/m; 1/3 K/m; 0 K/m. The inclined lines in Figure~\ref{Fig3} are related to the first saturation mechanism due to the particle depletion effect. The horizontal line corresponds to the second saturation mechanism caused by the momentum coupling of particles and turbulent flow with the mass loading parameter $m_p n_{\rm cl}/ \rho_g \approx 1/2$.
In Fig.~\ref{Fig3} the following parameters (see below) have been used: ${\rm Re}=10^3$, $\nu=0.2$ cm$^2$/s, $\sigma_{_{T0}} =1/2$, and $c_{\rm s}=450$ m/s.

	Using Eqs.~(\ref{A15})--(\ref{A18}), and taking into account that $\beta=\meankappa \, \ell_D \ll 1$, we determine the integrals  $J_1$ and $J_2$:
\begin{eqnarray}
J_1=J_2 &=& \left({n_{\rm cl} \over
\meanN}\right)^2 \, \left[1 + {\mu-1 \over (\mu-1)^2 + \alpha^2}\right] .
\label{A19}
\end{eqnarray}
Since the main contributions to the integrals $J_1$ and $J_2$ are from the size
of the cluster, $\ell_D =a \, \ell_\eta / {\rm Sc}^{1/2} \propto \nu^{3/4}$, these integrals depend on the kinematic viscosity $\nu$.

\subsection{Effective penetration length of radiation}
	
Substituting Eq.~(\ref{A19}) into Eq.~(\ref{A13}) for the effective absorption coefficient, we obtain the following expression for the effective penetration length of radiation,  $L_{\rm eff} \equiv 1/\kappa_{\rm eff}$:
\begin{eqnarray}
{L_{\rm eff} \over L_a} =1 + {2 a \over {\rm Sc}^{1/2}} \, \left({n_{\rm cl} \over \meanN}\right)^{2} \, \left({\ell_\eta \over L_a}\right)\, \left[1 + {\mu-1 \over (\mu-1)^2 + \alpha^2}\right]  .
\nonumber\\
\label{A20}
\end{eqnarray}

Let us analyse the obtained result for the effect of turbulent particle clustering on the increase of the effective penetration length of radiation $L_{\rm eff}/L_{a}$.
There are two physical effects related to particle clustering which affect the radiative transfer: (i) formation of the transparent for radiation
windows between particle clusters, and
(ii) screening of particles inside the clusters from the radiation for optically thick clusters.
The analysis performed in the present study yields the effective absorption coefficient, $\kappa_{\rm eff}$, that takes into account the collective effects of turbulent clusters
on the radiation transfer.
This effect determines the formation of the transparent windows between particle clusters.
On the other hand, the presented analysis cannot take into account
the screening effect for optically thick clusters.
Therefore, the obtained results  give a lower limit for the increase in  the penetration  length of radiation, $L_{\rm eff}$, whereas the overall effect may be much stronger.

It should be noted that in \cite{Frankel(2017),Frankel-PhD-Thesis(2017)}  authors attempted, following our recent paper \cite{Liberman(2017)}, to calculate the increase in the radiation transmission through gas-particle mixtures caused by the particle clustering for an isothermal turbulence. However, the analytical expression for the ratio $L_{\rm eff}/L_{a}$ obtained in \cite{Frankel(2017)} differs from Eq.~(18) in \cite{Liberman(2017)} or Eq.~(\ref{A20}) in the present paper. The expression for the particle radial distribution function used in \cite{Frankel(2017)} for the derivation of the ratio $L_{\rm eff}/L_{a}$  corresponds to weak clustering regime when the clustering instability is not excited.
%IR:
%Also the derivation of the ratio $L_{\rm eff}/L_{a}$ based on Eq.~(16) in %\cite{Frankel(2017)}, is incorrect.
The derivation of the ratio $L_{\rm eff}/L_{a}$ performed in \cite{Frankel-PhD-Thesis(2017)}
is based on the assumption of a normally-distributed optical path length.
It is also assumed very small particle clusters.
%IR.
As regarding the numerical results in \cite{Frankel(2017)}, for the used Reynolds number (based on the Taylor microscale) less than 50 implies that turbulence in this case is not fully developed, so that the increase in $L_{\rm eff}$ in comparison with the length $L_{a}$ should be very small.

\begin{figure}
\vspace*{1mm} \centering
\includegraphics[width=11.0cm]{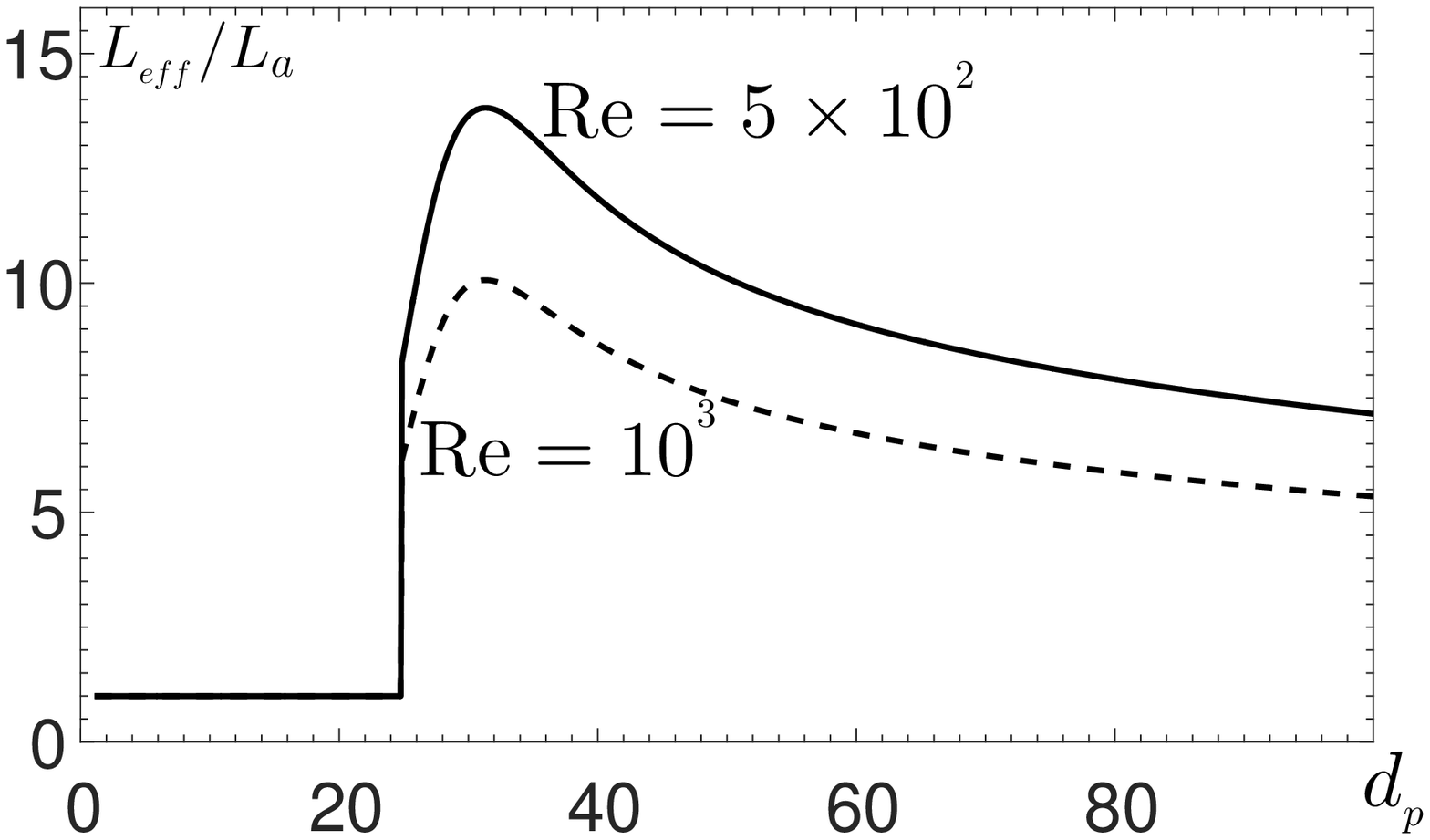}
\includegraphics[width=11.0cm]{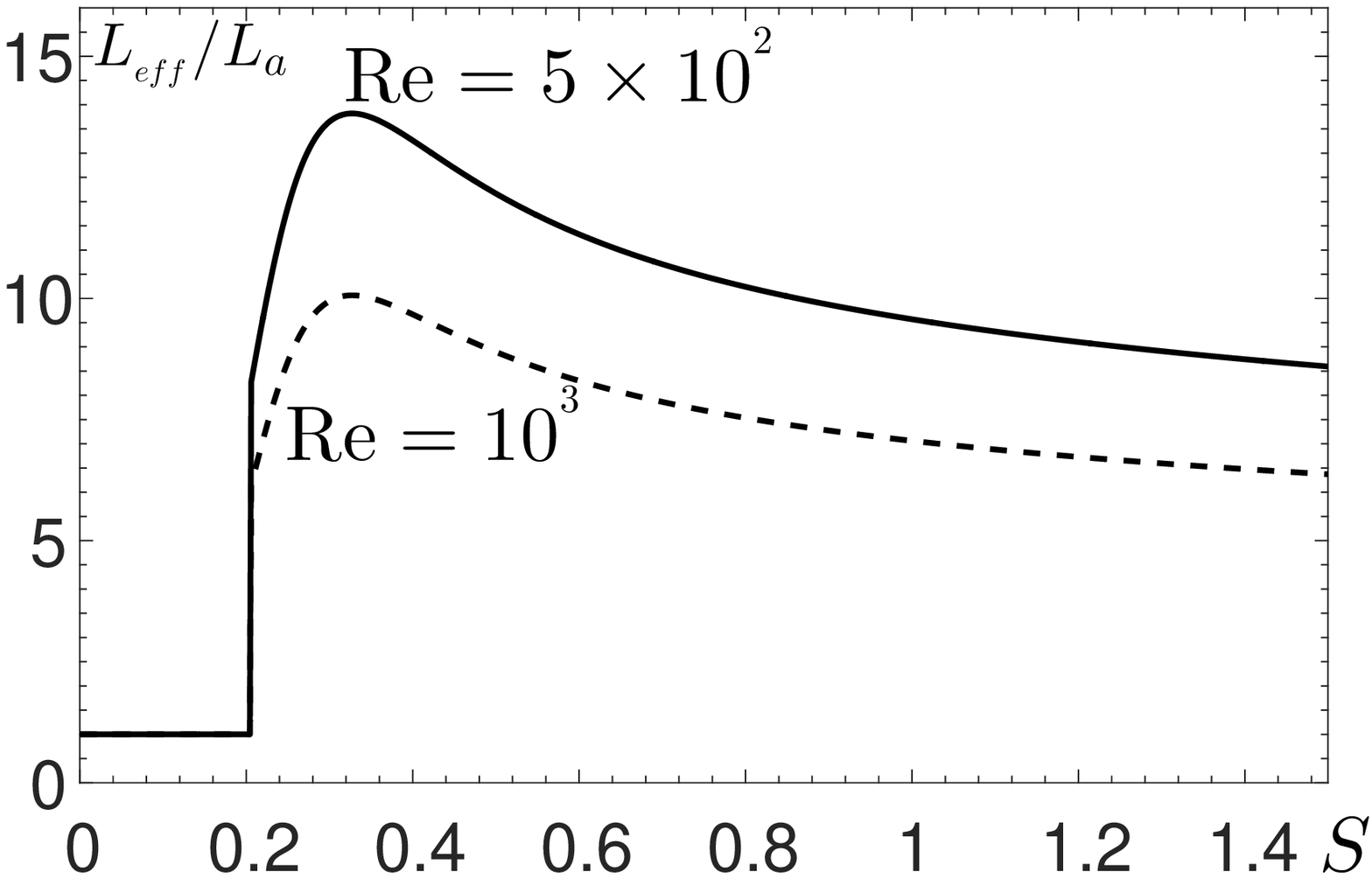}
\caption{\label{Fig4}
The ratio $L_{\rm eff}/L_{a}$ versus the particle sizes (top panel) and the Stokes number (bottom panel) for the particle clustering in isothermal turbulence with different Reynolds numbers: Re= $5 \times 10^2$ (solid); $10^3$ (dashed).}
\end{figure}

We first consider turbulence in the absence of the mean temperature gradient (isothermal turbulence).
For isothermal turbulence the particle clustering instability occurs only when
the particle diameter $d_p>25 \mu$m (see \cite{Elperin(2002),Elperin(2007),Elperin(2013)}).
Figure~\ref{Fig4} shows the ratio of $L_{\rm eff}/L_{a}$  versus the particle sizes (Figure~\ref{Fig4}, top panel) and versus the Stokes number (Figure~\ref{Fig4}, bottom panel) determined for the case of particle clustering in turbulence with different Reynolds numbers. The following parameters have been used for methane-air at normal conditions: $\nu=0.2$ cm$^2$/s,  $c_{\rm s}=450$ m/s, $n_{\rm cl} /\meanN = 500$ and $\sigma_{_{T0}} =1/2$. Figure~\ref{Fig4} demonstrates that increase of the effective penetration length of radiation in isothermal turbulence is more than one order of magnitude for the particles with $d_p>25 \mu$m.

As follows from Eq.~(\ref{A20}) the ratio $L_{\rm eff}/L_{a}$ strongly dependent on $\left(n_{\rm cl} /\meanN\right)^{2}$.
It was shown in \citep{Elperin(2013),Elperin(2015)} that for temperature stratified turbulence
the ratio $n_{\rm cl} / \meanN$ can increase up to three orders of magnitude.
However, in the case of isothermal turbulence the rate of the particle clustering is rather low. It is more than order of magnitude less than the rate of the particle clustering for a temperature stratified turbulence (see Figure~1 in \cite{Elperin(2013)}). Therefore, for isothermal turbulence the density of particles in clusters is unlikely to increase significantly. In this case, clusters either do not survive for a long time, or they are absorbed by the advancing flame long before the density of particles in clusters can increase significantly. It was shown in \citep{Elperin(2013),Elperin(2015)} that for temperature stratified turbulence the ratio $n_{\rm cl} /\meanN$  can increase up to three orders of magnitude.

\begin{figure}
\vspace*{1mm} \centering
\includegraphics[width=11.0cm]{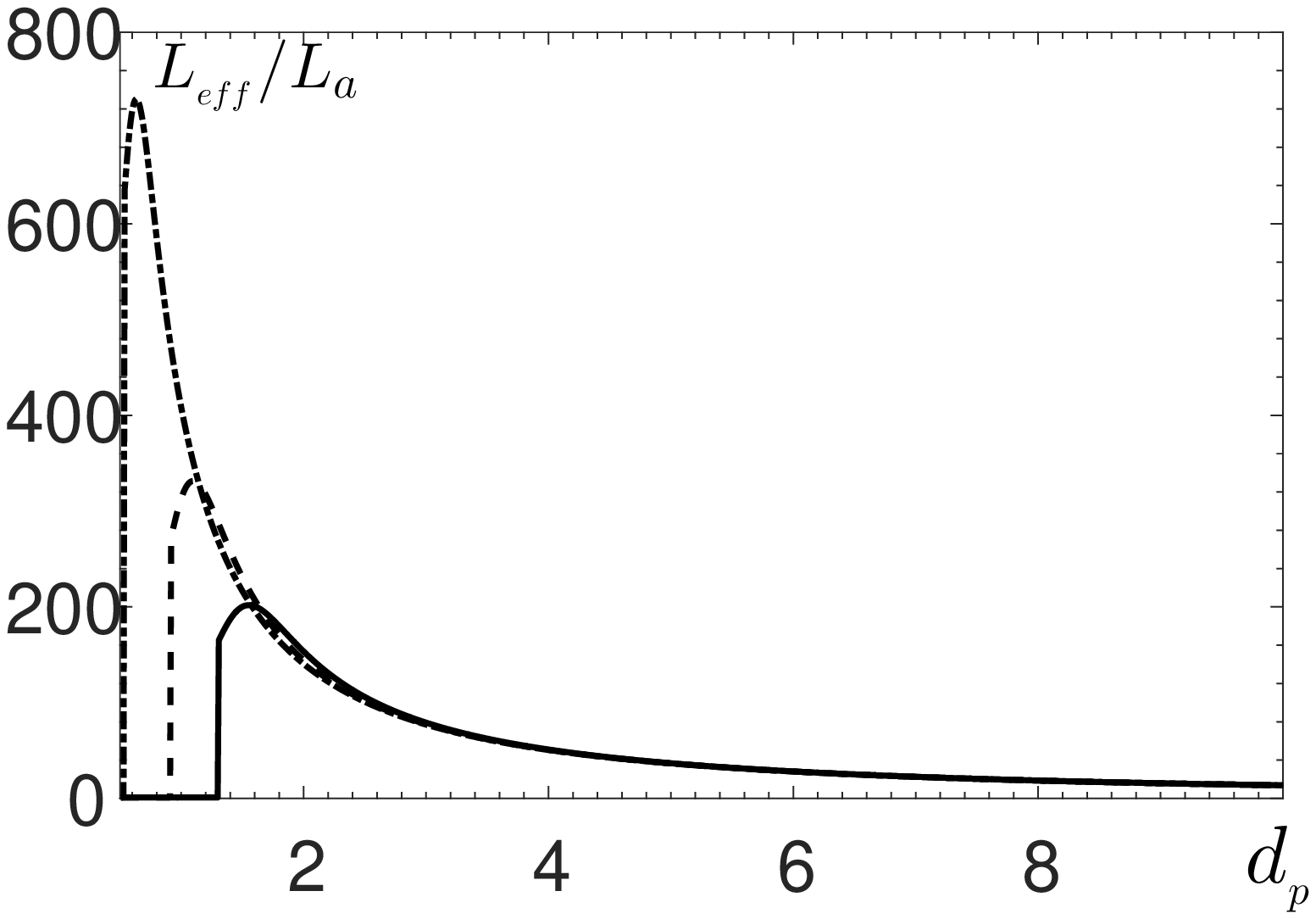}
\includegraphics[width=11.0cm]{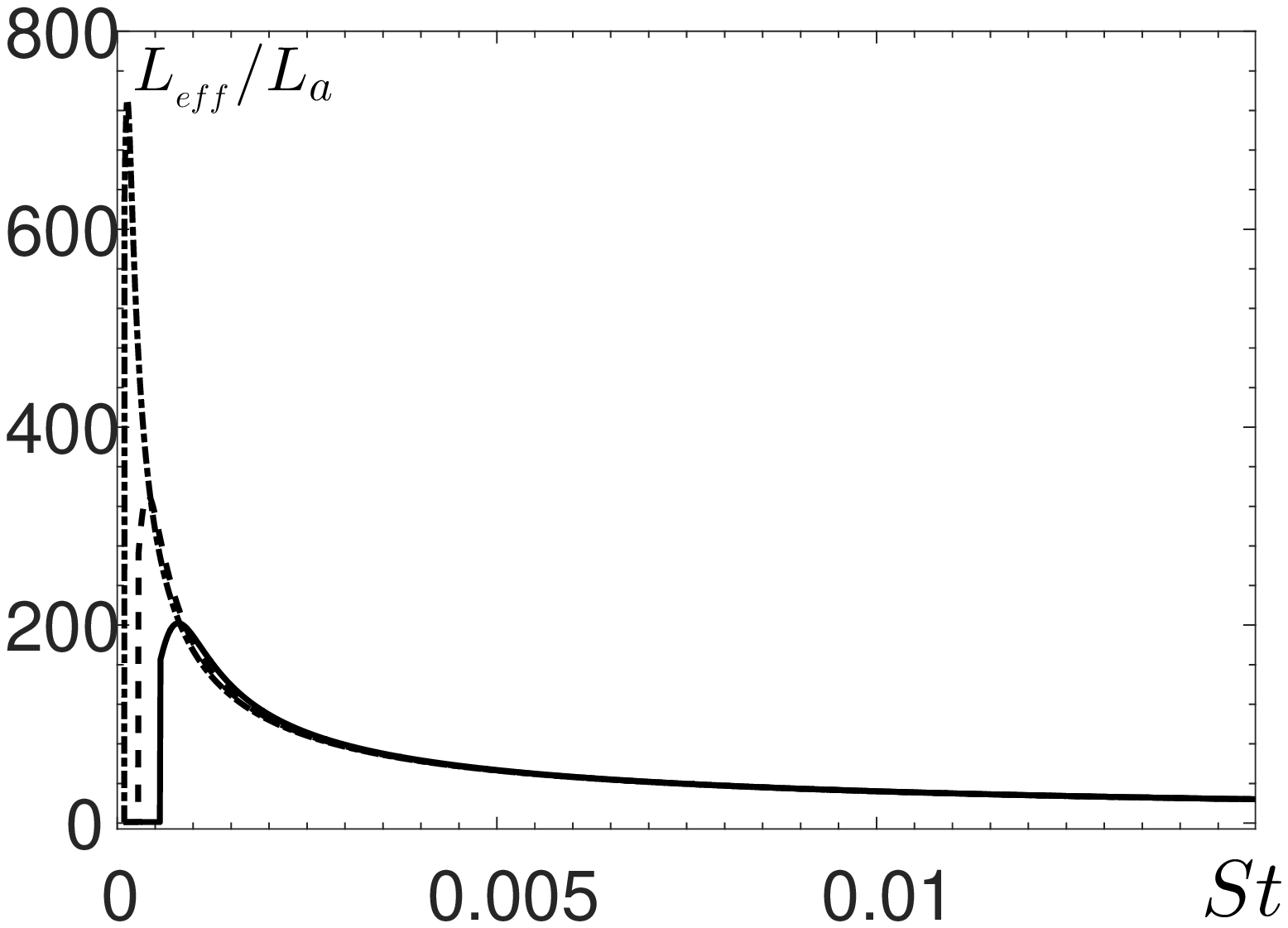}
\caption{\label{Fig5}
The ratio $L_{\rm eff}/L_{a}$ versus the particle sizes (top panel) and the Stokes number (bottom panel) for different mean temperature gradients $|{\bm \nabla} \meanT|=$ 0.5 K/m (solid) , 1 K/m (dashed), 3 K/m (dashed-dotted). The particle diameter is in microns.}
\end{figure}

There are many reasons for the formation of an inhomogeneity in the temperature distribution in the flow ahead of the front of the advancing flame.  Therefore, it is natural to assume that in dust explosions the fluid temperature in the flow ahead of the advancing primary flame is non-uniform.
As a representative example, we choose the mean temperature gradient
in the range of 0.5 -- 3 K/m.
In Figure~\ref{Fig5} we show the ratio $L_{\rm eff}/L_{a}$ versus the particle sizes (top panel) and the Stokes number (bottom panel) for different values of the mean temperature gradients.
In this figure the following parameters typical for the turbulent flow in unconfined dust explosion have been used: ${\rm Re}=5 \times 10^4$, $\ell_0=1$ m, $u_0=1$ m/s; $\nu=0.2$ cm$^2$/s, $n_{\rm cl} /\meanN = 500$, $\sigma_{_{T0}} =1/2$, and $c_{\rm s}=450$ m/s for methane-air at normal conditions.
As follows from Figure~\ref{Fig5}, the increase of the effective radiation
penetration length for the temperature stratified turbulence
is much stronger than that for isothermal turbulence
($L_{\rm eff}$ increases up to three orders of magnitude compared to $L_{a}$).

\section{Radiation-induced secondary explosions}

At the early stage of dust explosions, the accidentally ignited subsonic flame accelerates due to hydrodynamic instabilities and interaction with the turbulent flow. The pressure waves, produced by the accelerating flame, run away and produce turbulence ahead of the advancing flame. With the increase of the primary flame surface and the flame velocity, the parameters of the turbulent flow ahead of the advancing flame, ${\rm Re}$, $\ell_0$, $u_0$, St and ${\bm \nabla} \meanT$ change continuously. The dust particles in the generated turbulent flow assemble in clusters during time of the order of milliseconds.

Analysis performed in the previous section shows that clustering of particles in the temperature stratified turbulent flow ahead of the primary flame may increase the radiation penetration length up to 2--3 orders of magnitude (see Figure~\ref{Fig5}).
This effect ensures that clusters of particles, are exposed to and heated by the radiation from the primary flame for sufficiently long time to become ignition kernels in a large volume ahead of the flame. The multi-point radiation-induced ignition of the surrounding fuel-air increases effectively the total flame area, so the distance, which each flame has to cover for a complete burn-out of the fuel, is substantially reduced. It results in a strong increase of the effective combustion speed, defined as the rate of reactant consumption of a given volume, and overpressures, required to account for the observed level of damages in unconfined dust explosions. If, for example, the radiation absorption length of evenly dispersed particles with spatial mass density 0.03 kg/m$^3$ was in the range of a few centimeters, dust particles assembled in the
clusters of particles, are sufficiently heated by radiation at distances up to 10--20 m ahead of the advancing flame.

The experimentally measured \citep{Beyrau(2013),Beyrau(2015)} ignition time of the fuel-air by the radiatively heated particles is 100 ms for 10 $\mu$m nonreactive particles.
The level of thermal radiation of $S \approx 300 - 400$ kW/m$^2$ \citep{Hadjipanayis(2015)}, emanating from hot combustion products in dust explosions, is sufficient to rise the temperature of particles by   $\Delta T_p \approx 1000$~K during ${\tau}_{\rm heating}= (\rho_p \, d_p \, c_{\rm p} \,
\Delta T_p)/2S < 10$ ms, where $c_{\rm p}$ is the particle specific heat.
The interphase (particle-gas) energy exchange time is
$\tau_{pg} = \rho_p \, d_p^2 \, c_{\rm p} / 6 \lambda {\rm Nu} < 1$ ms, where $\lambda= {\chi} {\rho_g}c_{\rm p} $ is the gas thermal conductivity, $ \chi$ is the thermal diffusivity, $c_{\rm p}$ is the heat capacity of the gas, and ${\rm Nu} \approx 2$ is the Nusselt number \citep{Acrivos(1962)}. It is known that in order for ignition to occur, a gas volume with a size that is of the order of the typical flame thickness, $r_f$ (for CH$_4$/air ${r_f} \approx 2$mm), has to be heated to ignition temperature to ignite a self-sustained combustion in a gas mixture. This means that the time it takes for an isolated particle to heat the surrounding gas to a level where it can ignite is relatively long  $(t_\chi \sim r^2/\chi \approx 500$ ms).
For typical parameters of dust explosions, the Kolmogorov viscous time scale is $\tau_{\eta} \sim 5 \times 10^{-3}$ s, and of the same order is the life time of turbulent clusters.
Since the life-time of clusters is larger than the ignition time,
the theory presented in previous section describes satisfactory the origin and mechanism of the secondary explosion.

Numerical simulations of fuel-air ignition by a radiatively heated cluster of particles, show that the ignition time-scales decrease approximately as  $1/n_{\rm cl}^{1/3}$, with the increase of the particle number density $n_{\rm cl}$  in the cluster. In this case the cluster of radiatively heated particles ignites the fuel-air mixture in a much shorter time, since the gas mixture is heated by thermal conduction within a small distances,  $\Delta r = 1/n_{\rm cl}^{1/3}$, corresponding to the particle separation in the cluster.

The above analysis indicates that the ignition time of the mixture by the radiatively heated cluster of particles, is of the order of 10 ms, the effective propagating rate of combustion caused by the secondary explosion, can be estimated as $L_{\rm eff} / \tau_{\rm ign} \sim 10^3$ m \, s$^{-1}$. Such rate of the secondary explosion corresponds to the intensity of a shock wave with Mach number, ${\rm Ma}=2 - 3$, producing overpressures of 8--10 atm. This may explain the mechanism of the secondary explosion and the level of damages observed in unconfined dust explosions.

\begin{figure}
\vspace*{1mm} \centering
\includegraphics[width=11.0cm]{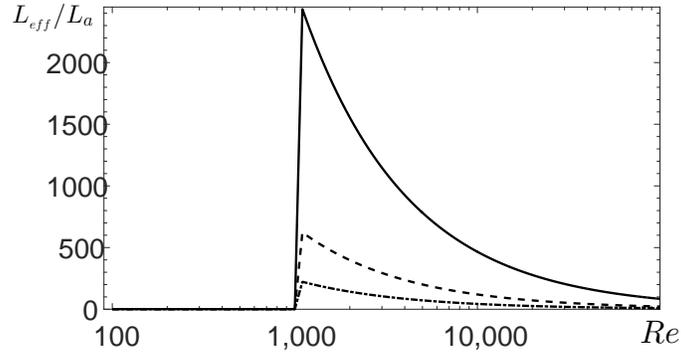}
\caption{\label{Fig6}
The ratio $L_{\rm eff}/L_{a}$ versus the Reynolds number ${\rm Re}$ for different particle diameter $d_p=$ 1 $\mu$m (solid), 5 $\mu$m (dashed), 10 $\mu$m (dashed-dotted),
and for the temperature gradients $|{\bm \nabla} \meanT|= 3$ K/m.}
\end{figure}

\begin{figure}
\vspace*{1mm} \centering
\includegraphics[width=11.0cm]{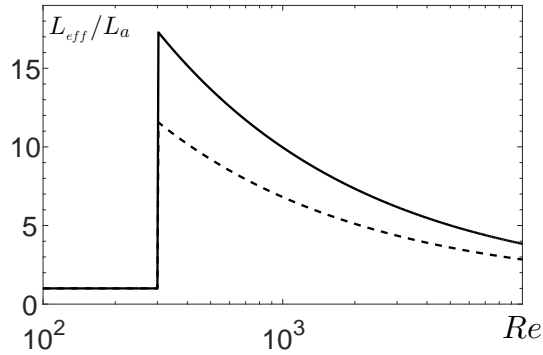}
\caption{\label{Fig7}
The ratio $L_{\rm eff}/L_{a}$ for isothermal turbulence versus the Reynolds number ${\rm Re}$ for different particle diameter $d_p=$ 31.3 $\mu$m (solid), 40 $\mu$m (dashed).}
\end{figure}

	Figure~\ref{Fig6} shows the function $L_{\rm eff}/L_{a}$ versus Reynolds numbers ${\rm Re}$ calculated for particles of different diameters. It is seen that a significant increase of the radiation penetration length caused by the particle clustering occurs within a rather narrow interval of turbulent parameters. The effect is much weaker if the flow parameters ahead of the flame front are changed and appear out of the ''range of transparency''. Such dependence of   $L_{\rm eff}/L_{a}$  versus Reynolds numbers suggests possible explanation of the episodic nature of explosion in the Buncefield incident described in \cite{Atkinson(2011)}.
For comparison, in Figure~\ref{Fig7} we show the ratio $L_{\rm eff}/L_{a}$ versus Reynolds numbers ${\rm Re}$ for isothermal turbulence. As we mentioned before the effect for the
isothermal turbulence is much weaker than that for the stratified turbulence.

According to the analysis by Atkinson and Cusco of the Buncefield explosion \citep{Atkinson(2011)}: ''The high overpressures in the cloud and low average rate of flame advance can be reconciled if the rate of flame advance was episodic, with periods of very rapid combustion being punctuated by pauses when the flame advanced very slowly. The widespread high overpressures were caused by the rapid phases of combustion; the low average speed of advance was caused by the pauses.'' From the beginning parameters of the turbulent flow ahead of the advancing flame vary continuously and finally fall within the ''range of transparency'', when the radiation penetration length increased considerably (Figures~\ref{Fig6} and~\ref{Fig7}).

Since the primary flame is a deflagration, propagating with velocity 
of the order of a few meters per second, 
the duration of this stage is the longest timescale in the problem. During this time the particle clusters ahead of the flame are exposed to and heated by the forward radiation for a sufficiently long time to become ignition kernels in a large volume ahead of the flame initiating the secondary explosion. Since the parameters of the turbulent flow are changed after the secondary explosion, the rapid phase of combustion is interrupted until the shock waves produced by the secondary explosion dissipate. The next phase continues until the parameters of turbulence in the flow ahead of the combustion wave will fall again within the interval corresponding to the ''transparent window'', such that the increased ratio $L_{\rm eff}/L_{a}$, caused by the particle clustering, provides conditions for the next secondary explosion.

\section{Conclusions}
\label{sec:conclusions}

The present study has shown that the mechanism of the secondary explosion in unconfined dust explosions and large vapor cloud explosions can be explained by the turbulent clustering of dust particles. The latter include the effect of considerable increase of the radiation penetration length, the formation of ignition kernels into the turbulent flow caused by the primary flame, and the subsequent formation of the secondary explosions, which is caused by the impact of forward thermal radiation. The mechanism of multi-point radiation-induced ignitions due to the turbulent clustering of particles ensures that ignition of the gas mixture by the radiatively-heated clusters occurs rapidly and within a large volume ahead of the primary flame.

The described picture of the unsteady combustion consisting of the rapid combustion producing high overpressures, which is punctuated by subsequent slow combustion, is well consistent with  \cite{Atkinson(2011)} analysis of the Buncefield explosion. Although details of the real physical processes can be different, the proposed theoretical model describes the observed in \cite{Atkinson(2011)} episodic nature of combustion in dust explosions, and more importantly it captures the most important relevant physics. Finally, the obtained analytical solution can serve as a benchmark for numerical simulations of dust explosions, which do not need to rely on the simplifying assumptions.

	The secondary explosion acts as an accelerating piston producing a strong pressure wave, which steepens in a shock wave. The intensity of the associated shock wave depends on the rate of progress of the secondary explosion and can be determined by using numerical simulations. It is possible that for a relatively large dust particles the condition, $\tau_{\rm ign} \, c_{\rm s} \ll L_{\rm eff}$, does not hold. Nevertheless, the increase of the effective penetration length of radiation gives rise to the secondary explosion, which propagates with the velocity  $U_{\rm exp} \approx L_{\rm eff}/\tau_{\rm ign}$, which is considerably larger than the velocity of the primary flame. More detailed analysis, i.e. taking into account the particle size distribution, the inter-particle collisions, possible coalescence of particles inside the clusters, and the effect of the gravitational sedimentation of large particles, can be only done using numerical simulations.

Finally it should be mentioned that transition to detonation, for example, supported by the forward thermal radiation in some circumstances cannot be ruled out. It should be emphasized that the effect of the strong increase of the radiation penetration length due to turbulent clustering of particles goes beyond the dust explosion applications and has many implications in astrophysical and atmospheric turbulence \citep{Elperin(2015),Owocki(2016),Sundqvist(2012)}.

\section*{Acknowledgement(s)}
Helpful discussions with G. Sivashinsky, A. Mani and
participants of the NORDITA program on “Turbulent Combustion” 
are gratefully acknowledged.

\section*{Funding}
The authors are grateful to the Research Council of Norway under the FRINATEK (grant No. 231444) for providing financial support for the current study. M.L. gratefully acknowledges support of this work provided by the opening project number KFJJ17-08M of State Key Laboratory of Explosion Science and Technology, Beijing Institute of Technology.

\end{document}